\documentclass[12pt,a4paper]{article}
\newcommand{\be}{\begin{equation}}
\newcommand{\en}{\end{equation}}
\newcommand{\bea}{\begin{eqnarray}}
\newcommand{\ena}{\end{eqnarray}}

\def\be {\begin{equation}}
\def\ee {\end{equation}}
\def\bea {\begin{eqnarray}}
\def\eea {\end{eqnarray}}
\def\bc {\begin{center}}
\def\ec {\end{center}}
\def\bfg {\begin{figure}}
\def\efg {\end{figure}}
\def\bi {\begin{itemize}}
\def\ei {\end{itemize}}

\oddsidemargin -.3cm \textwidth 16.5cm \textheight 22cm  \voffset -1cm \topmargin 1cm \fontsize{12pt}{14pt}\selectfont
\begin{document}

\title{\textbf{Regular black holes with a nonlinear electrodynamics source}}

\author{Leonardo Balart$^{1*}$ and Elias C. Vagenas$^{2+}$ \\    \\$^{1}$ \small Departamento de Ciencias F\'{\i}sicas, 
Facultad de Ingenier\'{\i}a y Ciencias, \\ \small Universidad de La Frontera, Casilla 54-D, Temuco, Chile and \\ \\
$^{2}$ \small Theoretical Physics Group, Department of Physics, \\ 
\small Kuwait University, P.O. Box 5969, Safat 13060, Kuwait \\
\emph{\small $^{*}$email:leonardo.balart@ufrontera.cl} \\ \emph{\small $^{+}$email: elias.vagenas@ku.edu.kw}
}

\date{}

\maketitle

\begin{abstract}
\par\noindent
We construct several charged regular black hole metrics employing mass distribution functions
which are inspired by continuous probability distributions. Some of these metrics satisfy the weak energy condition and
asymptotically behave as the Reissner--Nordstr\"{o}m black hole. In each case, the source to the Einstein equations corresponds
to a nonlinear electrodynamics model, which in the weak field limit becomes the Maxwell theory (compatible with the Maxwell weak
field limit or approximation). Furthermore, we include other regular black hole solutions that satisfy 
the weak energy condition and some of them correspond to the Maxwell theory in the weak field limit.

\end{abstract}

\section{Introduction}
\label{intro}

Charged regular black holes are solutions of Einstein equations that have horizons and, contrary to Reissner--Nordstr\"{o}m black holes which have singularities at the origin, their metrics as well as their curvature invariants $R$, $R_{\mu\nu} R^{\mu\nu}$, $R_{\kappa\lambda\mu\nu} R^{\kappa\lambda\mu\nu}$ are regular everywhere~\cite{Ansoldi:2008jw}. This type of black holes violates the strong energy condition somewhere in the spacetime (see, e.g.,
Ref.~\cite{Elizalde:2002yz} or Ref.~\cite{Zaslavskii:2010qz}); however, some of these solutions satisfy the weak energy condition
(WEC) everywhere.
Those that satisfy the WEC necessarily have a de Sitter center~\cite{Dymnikova:2001fb}. In addition, there are other 
features that characterize regular black holes which are due to the nonlinearities of the field equations. 
For instance, the thermodynamic quantities of these black
holes do not satisfy the Smarr formula (see, e.g., Ref.~\cite{Breton:2004qa}), the identity of 
Bose--Dadhich~\cite{Bose:1998uu} refers to the relation between the Brown--York energy, and the Komar charge 
is not satisfied by regular black holes~\cite{Balart:2009xr}.

Several regular black hole solutions have been found by coupling gravity to nonlinear electrodynamics theories
(for a more detailed review see Ref.~\cite{Ansoldi:2008jw} and references cited therein). Two of these solutions satisfy the WEC
and asymptotically behave as the Reissner--Nordstr\"{o}m black hole; one of them was reported by Ay\'on-Beato and
Garc\'{\i}a in Ref.~\cite{AyonBeato:1998ub} (see also Ref.~\cite{AyonBeato:2004ih} which is a generalization of this case) and
the other one was presented by Dymnikova in Ref.~\cite{Dymnikova:2004zc}. Additionally Ay\'on-Beato and Garc\'{\i}a obtained other
solutions in Refs.~\cite{AyonBeato:1999rg} and \cite{AyonBeato:1999ec} (see also Ref.~\cite{Bronnikov:2000vy}), which asymptotically
behave as the Reissner--Nordstr\"{o}m solution but do not satisfy the WEC. These authors also considered the Bardeen regular black hole
\cite{Bardeen:1968}, which satisfies the WEC but asymptotically does not behave as the Reissner--Nordstr\"{o}m
solution, and interpreted it as gravity coupled to a theory of nonlinear electrodynamics for a self-gravitating magnetic monopole in
Ref.~\cite{AyonBeato:2000zs}. Recently, the authors of this paper have presented a family of regular black hole metrics in 
Ref.~\cite{Balart:2014jia}, which by construction satisfy the WEC.
Other black hole solutions that satisfy the WEC have been found by considering a Gaussian
distribution for the mass density in Ref.~\cite{Nicolini:2005vd} and also for both the mass density and the charge density in
Ref.~\cite{Ansoldi:2006vg}. A noncharged black hole solution which is also regular and satisfies WEC  
was given by Dymnikova in Ref.~\cite{Dymnikova:1992ux}. A black hole solution similar to the Bardeen solution, 
i.e., one that satisfies the WEC but asymptotically does not behave as the Reissner--Nordstr\"{o}m 
solution, is given by Hayward in Ref.~\cite{Hayward:2005gi}. 
Other solutions of regular black holes have been obtained by the junction of two spherically symmetric regions in
Refs.~\cite{Elizalde:2002yz} and~\cite{Lemos:2011dq}. In particular, Ref.~\cite{Elizalde:2002yz} carried
out a detailed analysis of the energy conditions of these solutions, and the authors showed that not all cases verify the WEC everywhere.
The solutions considered in Ref.~\cite{Elizalde:2002yz} satisfy the WEC, where the outer and the inner regions
are separated by an electrically charged spherically symmetric coat and the outer region behaves as a Reissner--Nordstr\"{o}m 
solution.

Using some of these examples, several articles have discussed some other features of this family of solutions 
or the particularities that differentiate this family of solutions with respect to Reissner--Nordstr\"{o}m solutions.
In Ref.~\cite{Eiroa:2010wm}, the Bardeen black hole was considered as a  gravitational lens, and the results were  compared with those obtained by a solution of Schwarzschild type.
The thermodynamics and evaporation process of the solution proposed by Hayward are discussed in Ref.~\cite{Myung:2007qt}.
The solution given in Ref.~\cite{AyonBeato:1998ub} has been considered with a cosmological constant in 
Ref.~\cite{Mo:2006tb}, both the thermodynamics and stability have been analyzed. 
There are various studies which consider the regular black hole defined by Ref.~\cite{AyonBeato:1999rg}. In particular,
the thermodynamics of this solution was studied in Ref.~\cite{Myung:2007av}, and the entropy of the corresponding extremal case was studied in Refs.~\cite{Matyjasek:2004gh}, \cite{Myung:2007xd}, and \cite{Myung:2007an}. It was also used to
construct a regular black hole solution in an asymptotically de Sitter universe in Ref.~\cite{Matyjasek:2008kn}. Likewise it was used for construction of the renormalized stress-energy tensor for the massive scalar fields case~\cite{Matyjasek:2000iy,Berej:2002xd} and for the
massive spinor and vector fields cases~\cite{Matyjasek:2007zz}. It was also used in the context of quadratic gravity in Refs.~\cite{Berej:2006cc} and~\cite{Matyjasek:2008yq}. Other studies have used two or more of the solutions of 
regular black holes mentioned in the previous paragraph: for the construction of gravastar models in Ref.~\cite{Lobo:2006xt} and for obtaining solutions
in $f(R)$ modified theories of gravity in Ref.~\cite{Hollenstein:2008hp}. The stability of some solutions  was  studied in
Refs.~\cite{Moreno:2002gg},~\cite{Breton:2005ye}, and~\cite{Breton:2014nba}.

In this work we construct several charged regular black hole metrics in the context of theories with nonlinear electrodynamics 
coupled to general relativity. The mass functions are inspired by continuous probability distributions. 
Some of these metrics satisfy the WEC and asymptotically behave as the Reissner--Nordstr\"{o}m black hole.

This paper is organized as follows. In Sec. 2, we construct a general regular black hole metric for mass distribution functions
that are inspired by continuous probability distributions. We also construct the corresponding electric field for each black hole solution
in terms of a general mass distribution function. In Sec. 3, we present two examples of black hole solutions 
employing the methodology developed in Sec. 2. In Sec. 4, we generalize the construction presented in Sec. 2 
by considering distribution functions raised to the power of a real number greater than zero. Moreover, we give two 
examples of black hole solutions by using this generalized construction. In Sec. 5, we consider a
particular distribution  in order to obtain a regular black hole metric, and then find the conditions under which the WEC is satisfied. 
In Sec. 6, we present some more regular black hole metrics that satisfy the WEC and, with the exception of one case, 
in the weak field limit the corresponding black hole solutions do not describe the Maxwell theory . In Sec. 7, we present 
regular black hole metrics that asymptotically behave as the Reissner--Nordstr\"{o}m solution and some satisfy the WEC. 
Finally, in Sec. 8, we briefly summarize our results. In Appendix A, we give some probability distributions, and in 
Appendix B, we give a brief description of the dual P formalism.

\section{General considerations}
\label{sec:2}

We consider the  line element for the most general static and spherically symmetric metric,

\begin{equation}
ds^2= -f(r) dt^2 + f(r)^{-1}dr^2 + r^2 (d \theta^2+\sin^2\theta d\phi^2)~,
\,\,\label{elem-gral} \, 
\end{equation}

\par\noindent
with

\begin{equation}
f(r) = 1 - \frac{2 m(r)}{r}
\,\,\label{metric-f} \, 
\end{equation}

\par\noindent
and where $m(r)$ is the mass function. The outer and the inner horizons are located at $r_+$ and $r_-$, respectively, 
satisfying $r_\pm = 2 m(r_\pm)$.

\par\noindent
To construct metrics of regular black holes where the invariant scalars and the electric fields are regular everywhere, 
we express the mass function $m(r)$ as 

\begin{equation}
m(r) = \, \frac{\sigma(r)}{\sigma_\infty }\,M \,\,\label{dens-mass} \,
\end{equation}

\par\noindent
where the distribution function $\sigma(r)$ satisfies $\sigma(r) > 0$ and $\sigma'(r) > 0$ for $r \geq 0$.  
Additionally, $\sigma(r)/r \rightarrow 0$ as $r \rightarrow 0$ and $\sigma_\infty = \sigma(r\rightarrow\infty)$ 
represents the normalization factor. We will employ distribution functions inspired by the shape of 
probability density functions that satisfy $\sigma(x) > 0$, $\sigma'(x)< 0$ for $x \geq 0$
and $\sigma'(x=0) \neq 0$ (see Appendix A). Moreover, depending on the choice of the distribution function, 
the variable $x$ is replaced in $\sigma(x)$ by $1/r$ or $1/\sqrt{r}$ with appropriate factors which depend
on the mass $M$ and the charge $q$ of the regular black hole.

\begin{table}
\begin{center}
\begin{tabular}{|c|}\hline
\textbf{Distribution function $\sigma(r)$}\\ \hline
$\exp\left(-\frac{q^2}{2 M r}\right)$ \\
\hline
$\left(\exp\left(\frac{q^2}{M r}\right) + 1\right)^{-1}$\\
\hline
$\exp\left(-\sqrt{\frac{2 q^2}{M r}}\right) / \left(1+\exp\left(-\sqrt{\frac{2 q^2}{M r}}\right)\right)^2$ \\
\hline
$2\left(\exp\left(\sqrt{\frac{q^2}{M r}}\right) + \exp\left(-\sqrt{\frac{q^2}{M r}}\right)\right)^{-1}$ \\
\hline
$\left(\frac{q^2}{M r}\right) / \left(\exp\left(\frac{q^2}{M r}\right) - 1\right)$ \\
\hline
$\left(\frac{6 q^2}{M r}\right) \exp\left(\sqrt{\frac{6 q^2}{M r}}\right) /  \left(\exp\left(\sqrt{\frac{6 q^2}{M r}}\right) - 1\right)^2$\\
\hline
\end{tabular}
\caption{Examples of probability distribution functions used in the mass functions of regular black holes.} \label{table1}
\end{center}
\end{table}

\par\noindent
The distribution functions are listed in Table 1 in the same order as in Appendix A and satisfy the condition

\begin{equation}
\frac{m(r)}{r} \rightarrow 0  \hspace{1cm}  \mbox{ when } \hspace{1cm} r \rightarrow 0  \,\,\label{limit-small} \,
\end{equation}
and asymptotically approach the Reissner--Nordstr\"{o}m metric
\begin{equation}
f(r) = 1 - \frac{2 m(r)}{r} \approx 1 - \frac{2 M}{r} + \frac{q^2}{r^2}\,\,\label{large} \, .
\end{equation}

\par\noindent
For each regular black hole solution of which the mass function is given by Eq.~(\ref{dens-mass}), we can obtain the electric field
by considering the components of the Einstein gravitational field equations $G_{\mu\nu} = 8 \pi T_{\mu\nu}$. Here,
the energy-momentum tensor $T_{\mu\nu}$ is  given as

\begin{equation}
T_{\mu\nu} = L(F) g_{\mu\nu} - L_F F_{\mu\alpha}F_\nu^{\,\, \alpha} \,\,\label{Tuv} \,
\end{equation}

\par\noindent
where the Lagrangian $L(F)$ depends on the Lorentz invariant $F = \frac{1}{4}F^{\mu\nu}F_{\mu\nu}$ 
and we have defined $L_F=dL/dF$.

\par\noindent
If we restrict to the electric field, i.e., we consider $F_{\mu\nu} = E(r) (\delta^0_\mu \delta^1_\nu - \delta^1_\mu \delta^0_\nu)$, and if
we employ Eqs.~(\ref{metric-f}) and~(\ref{dens-mass}) given here as well as Eq. (7) from Ref. \cite{Balart:2014jia}, 
then we can write the components of $G_{\mu\nu}$ as

\bea
G_{\,\, 0}^0 &=&  G_{\,\, 1}^1 = -\frac{2 M}{r^2} \frac{\sigma'(r)}{\sigma_\infty} = 8\pi \left(L(F) + E^2 L_F\right)
\label{G00} \\
G_{\,\, 2}^2  &=& G_{\,\, 3}^3 = -\frac{M}{r} \frac{\sigma''(r)}{\sigma_\infty}= 8\pi L(F)  
\label{G22} \, .
\eea

\par\noindent
In addition, the electromagnetic field equations $\nabla_\mu(F^{\mu\nu} L_F) = 0$ imply
\begin{equation}
E(r) L_F = - \frac{q}{4 \pi r^2}\,\,\label{elec} \, .
\end{equation}

\par\noindent
By subtracting Eq.~(\ref{G00}) from  Eq.~(\ref{G22}) and using Eq.~(\ref{elec}), we obtain a general expression for the
electric field $E$ of the regular black hole solutions 

\be
E(r) = -\frac{r^3}{2 q} \frac{M}{\sigma_\infty}\frac{d}{dr}\left(\frac{1}{r^2}\frac{d\sigma(r)}{dr}\right)\,\,\label{field.E} \, .
\ee

\par\noindent
After considering any one of the distribution functions listed in Table 1, the corresponding electric fields are regular everywhere 
and asymptotically behave as

\begin{equation}
E(r) \approx \frac{q}{r^2} \,\,\label{field.E.apr} \, ,
\end{equation}

\par\noindent
as expected from Eq.~(\ref{large}).

\par\noindent
It is important to note that  the distribution function $\sigma(r)$, which is associated with the mass function, converges to zero approximately
as $e^{-\frac{1}{r}}$ when $r \rightarrow 0$, and the same applies to the derivatives of any order of $\sigma(r)$.
Therefore, the curvature invariants, namely the Ricci scalar

\begin{equation}
R = \frac{2 M}{r^2 \sigma_\infty}(2 \sigma'(r) + r \sigma''(r)) \,\,\label{R-gral} \, ,
\end{equation}

\par\noindent
the  Ricci squared
\begin{equation}
R_{\mu\nu} R^{\mu\nu} = \frac{2 M^2}{r^4 \sigma_\infty^2}(4 \sigma'(r)^2 + r^2 \sigma''(r)^2) \,\,\label{ricci} \, ,
\end{equation}

\par\noindent
and the Kretschmann scalar
\begin{eqnarray}
R_{\kappa\lambda\mu\nu} R^{\kappa\lambda\mu\nu} &=&  \frac{4 M^2}{r^6 \sigma_\infty^2}[4 (3 \sigma(r)^2 - 4 r \sigma(r)\sigma'(r)
+ 2 r^2 \sigma'(r)^2)\nonumber \\
&& +\, 4 r^2(\sigma(r) - r \sigma'(r)) \sigma''(r) + r^4 \sigma''(r)^2] \,\,\label{krets} \, ,
\end{eqnarray}

\par\noindent
are regular everywhere.

\par\noindent
An equivalent description could be given by using the dual P formalism implemented in Ref.~\cite{Salazar:1987ap}
(some details of this formalism are presented in Appendix B).

\section{Two examples of regular black hole metrics}
\label{sec:3}

In this section, we consider two of the distribution functions listed in Table 1 in order to construct regular black hole metrics. 
\footnote{A similar treatment can also be applied to the rest of the distribution functions in Table 1.}

\par\noindent
First, we consider the exponential distribution, where the variable $x$ is replaced with $q^2/(2Mr)$ 
in Eq.~(\ref{exp-d}) of Appendix A (see also Table 1),
and we obtain

\begin{equation}
\sigma(r) = \exp\left(-\frac{q^2}{2 M r}\right)\,\,\label{dens-exp} \, 
\end{equation}

\par\noindent
where the normalization factor is

\begin{equation}
\sigma_\infty = 1 \,\,\label{norm-exp} \, .
\end{equation}

\par\noindent
Thus, the metric function is of the form

\begin{equation}
f(r) = 1 - \frac{2 M}{r} \,\, \exp\left(-\frac{q^2}{2 M r}\right)\,\,\label{func-exp} ,
\end{equation}

\par\noindent
which vanishes at the location of the horizons, i.e.,  $r_\pm$. By solving the resulting equation, namely  $f(r_{\pm})=0$, 
we find the  real roots for $r_\pm$,

\begin{equation}
r_+ = -\frac{q^2}{2 M \,\, W\hspace{-0.8ex}\left(0, -\frac{q^2}{4 M^2}\right)} \hspace{4ex}  \mbox{and} \hspace{4ex} 
r_- = -\frac{q^2}{2 M \,\, W\hspace{-0.8ex}\left(-1, - \frac{q^2}{4 M^2}\right)}
\,\,\label{sol-exp-2}\,\,\label{sol-exp},
\end{equation}

\par\noindent
where $W$ is Lambert's W function.

\par\noindent
The corresponding electric field can be directly obtained from Eq.~(\ref{field.E})

\begin{equation}
E(r) = \frac{q}{r^2}\left(1 - \frac{q^2}{8 M r}\right) \exp\left(-\frac{q^2}{2 M r}\right) \,\,\label{elec-exp} \,
\end{equation}

\par\noindent
which is regular everywhere and asymptotically behaves as $E(r) = q/r^2 + O(1/r^3)$. 
It should be noted that in this example  we have the extremal regular black hole when $|q| = 1.213 \, M$.

\par\noindent
Second, we consider the Fermi--Dirac-type distribution.
Thus, if $x$ is replaced with  $q^2/(M r)$ in Eq.~(\ref{ferm-dir}) of Appendix A (see also Table 1), 
we obtain the distribution function

\begin{equation}
\sigma(r) = \frac{1}{\exp(\frac{q^2}{M r})+1} \,\,\label{dens-fer} 
\end{equation}

\par\noindent
with normalization factor

\begin{equation}
\sigma_\infty = 1/2 \,\,\label{norm-fer} 
\end{equation}

\par\noindent
and the metric function is now written as

\begin{equation}
f(r) = 1 - \frac{2 M}{r}\left(\frac{2}{\exp(\frac{q^2}{M r})+1}\right)\,\,\label{func-fer} \, .
\end{equation}

\par\noindent
It is noteworthy  that this metric function corresponds to an Ay\'on-Beato and Garc\'{\i}a black hole~\cite{AyonBeato:1999rg}. 
In addition, this metric function satisfies the equation $f(r_\pm)=0$ for the location of the horizons, which now reads

\begin{equation}
r_\pm = 2 M \left(\frac{2}{\exp(\frac{q^2}{M r_\pm})+1}\right) \,\,\label{hor-fer} \, .
\end{equation}

\par\noindent
This equation has two real roots which again will be expressed in terms of  Lambert's W function~\cite{Matyjasek:2000iy}:

\begin{equation}
r_+ = - \frac{ 4 q^2}{4 M\,\,  W\hspace{-0.4ex}(0,-\frac{q^2}{4 M^2}\,e^{\frac{q^2}{4M^2}}) - \frac{q^2}{M}}   \hspace{4ex}  \mbox{and} \hspace{4ex} 
 r_- = - \frac{4 q^2}{4M\,\,  W\hspace{-0.4ex}(-1,-\frac{q^2}{4M^2}\,e^{\frac{q^2}{4 M^2}}) - \frac{q^2}{M}}\,\,\label{sol-fer} \, .
\end{equation}

\par\noindent
Employing Eq.~(\ref{field.E}), the corresponding electric field is now written as

\begin{equation}
E(r) = \frac{q}{r^2} \,\, \mbox{sech}^2\left(\frac{q^2}{2 M r}\right) \left(1 - \frac{q^2}{4 M r}\tanh\left(\frac{q^2}{2 M r}\right)\right)
\,\,\label{elec-fer} \, .
\end{equation}

\par\noindent
Furthermore, the corresponding extremal regular black hole is obtained when the value of the charge is $|q| = 1.055 \, M$ 
(see Refs.~\cite{Matyjasek:2004gh}, \cite{Myung:2007xd}, and \cite{Myung:2007an}).

\section{More regular black hole metrics}
\label{sec:4}

\par\noindent
The use of the distribution functions mentioned above in order to obtain regular black hole solutions can be extended 
by considering the metric function to take the form

\begin{equation}
f(r) = 1 - \frac{2 M}{r}\left(\frac{\sigma(\beta r)}{\sigma_\infty}\right)^\beta   \,\,\label{case-beta} ,
\end{equation}

\par\noindent
where the function $\sigma(\beta r)$ corresponds to any one of the mass functions listed in Table 1, 
but with an additional factor $\beta > 0$ for the variable $r$.

\par\noindent
In this case, we can see that the mass function satisfies the condition

\begin{equation}
m(r) = M \left(\frac{\sigma(\beta r)}{\sigma_\infty }\right)^\beta \rightarrow M  \hspace{1cm}  \mbox{when} 
\hspace{1cm} r \rightarrow \infty  \,\,\label{lim-inf-beta} 
\end{equation}

\par\noindent
in the same way as the distribution function satisfies
\begin{equation}
\frac{\sigma(r)}{\sigma_\infty } \rightarrow 1  \hspace{1cm}  \mbox{when} \hspace{1cm} r \rightarrow \infty  \,\,\label{lim-inf-1} \, .
\end{equation}

\par\noindent
Furthermore, using the same argument as above, one can find that all curvature invariants are regular.

\par\noindent
To illustrate these cases, let us consider here two examples. The first one is constructed 
by employing the Fermi--Dirac distribution function. In this case, the metric function is written as

\begin{equation}
f(r) = 1 - \frac{2 M}{r} \left[\frac{2}{\exp\left(\frac{q^2}{\beta M r}\right) + 1}\right]^\beta
\,\,\label{ferm-bet}  \, .
\end{equation}

\par\noindent
The outer and the inner horizons of this metric function can be found numerically for each value of $\beta$.
The corresponding electric field  can be obtained by using Eq.~(\ref{field.E}); thus

\begin{eqnarray}
E(r) &=& \frac{q}{r^2} \,\, \exp \left(\frac{(1-\beta)q^2}{2 \beta M r}\right)\left[\mbox{sech}\left(\frac{q^2}{2 \beta M r}\right)\right]^{1+\beta}
\nonumber \\
&\times& \left[1 - \frac{q^2}{4 M r}\tanh\left(\frac{q^2}{2 \beta M r}\right)+\frac{1}{4 \beta M r}\left(\frac{1-\beta}
{\exp\left(\frac{q^2}{\beta M r}\right)+1}\right)\right] \,\,\label{elec-ferm-bet} \, .
\end{eqnarray}

\par\noindent
If we set $\beta \rightarrow 0$ in Eq.~(\ref{ferm-bet}), then we obtain the metric

\begin{equation}
f(r) = 1 - \frac{2 M}{r}\exp\left(-\frac{q^2}{M r}\right)
   \,\,\label{ferm-bet-z} \, ,
\end{equation}

\par\noindent
while when we set $\beta \rightarrow \infty$, we reproduce the metric function given by Eq.~(\ref{func-exp}), i.e.,

\begin{equation}
f(r) = 1 - \frac{2 M}{r} \exp\left(-\frac{q^2}{2 M r}\right)
\,\,\label{ferm-bet-inf}  \, .
\end{equation}

\par\noindent
It should be noted the difference in a factor of $2$ between the exponents of Eqs.~(\ref{ferm-bet-z}) and~(\ref{ferm-bet-inf}).
In  Table 2, we  list some values of $\beta$ and the corresponding charges in order to construct 
the extremal regular black hole metric for this  example.

\begin{table}
\begin{center}
\begin{tabular}{|c|c|}\hline
$\beta $ & \textbf{Extremal case} \\ \hline
$0.5$ & $q \approx 0,991 \, M$\\
\hline
$0.7$ & $q \approx 1.023 \, M$\\
\hline
$1$ & $q \approx 1.055 \, M$\\
\hline
$2.4$ & $q \approx 1.124 \, M$\\
\hline
$4$ & $q \approx 1.153 \, M$\\
\hline
$7$ & $q \approx 1.175 \, M$\\
\hline
$10$ & $q \approx 1.186 \, M$\\
\hline
$100$ & $q \approx 1.210 \, M$\\
\hline
\end{tabular}
\caption{Values of $\beta$ and the corresponding charges for the case of an extremal regular black hole when we consider the 
metric function given by Eq. ~(\ref{ferm-bet}).} \label{table2}
\end{center}
\end{table}

\par\noindent
The second example is given by using the logistic distribution function for which the metric function reads

\begin{equation}
f(r) = 1 - \frac{2 M}{r} \left[\frac{4 \exp \left(-\sqrt{\frac{2 q^2}{\beta M r}}\right)}{\left(1 + 
\exp \left(-\sqrt{\frac{2 q^2}{\beta M r}}\right)\right)^2}\right]^\beta
\,\,\label{logistic-bet}  \, .
\end{equation}

\par\noindent
As before the horizons can be found numerically  for each value of $\beta$. Moreover, the corresponding 
electric field can be obtained using Eq.~(\ref{field.E}):

\begin{eqnarray}
E(r) &=& \frac{q}{r^2} \,\, \frac{1}{8} \left[\mbox{sech}\left(\sqrt{\frac{q^2}{2 \beta M r}}\right)\right]^{2(1+\beta)} \nonumber \\
&\times& \left[(1+\beta) - \beta \cosh\left(\sqrt{\frac{2 q^2}{\beta M r}}\right)+ 7 \, \sqrt{\frac{\beta M r}{2 q^2}}
\sinh\left(\sqrt{\frac{2 q^2}{\beta M r}}\right)\right]
\,\,\label{elec-logistic-bet} \, .
\end{eqnarray}

\par\noindent
At this point a number of comments is in order. First,  it is noteworthy that the above expression for the electric field as well as the electric field given by Eq.~(\ref{elec-ferm-bet}) 
asymptotically behave as $E=q/r^2$.
Second, if we set $\beta \rightarrow 0$ in the metric function given by Eq.~(\ref{logistic-bet}), then we recover the 
Schwarzschild black hole

\begin{equation}
f(r) = 1 - \frac{2 M}{r} \,\,\label{logistic-bet-z}  \, ,
\end{equation}

\par\noindent
while if we set $\beta \rightarrow \infty$, then we get the metric function given by Eq.~(\ref{func-exp}):

\begin{equation}
f(r) = 1 - \frac{2 M}{r} \exp\left(-\frac{q^2}{2 M r}\right)
\,\,\label{logistic-bet-inf}  \, .
\end{equation}

\par\noindent
Third, it should be stressed  that in the limit $\beta \rightarrow \infty$ all regular black hole metrics constructed by employing 
Eq.~(\ref{case-beta}) and the distribution functions listed in Table 1 satisfy the condition

\begin{equation}
1 - \frac{2 M}{r}\left(\frac{\sigma(\beta r)}{\sigma_\infty}\right)^\beta \,\,\, \rightarrow \,\,\, 1 - \frac{2 M}{r} \,\, 
\exp\left(-\frac{q^2}{2 M r}\right)\,\,\label{limit-beta} \,.
\end{equation}

\par\noindent
Finally, when we set $\beta \rightarrow 0$,  for some of the cases, we recover the Schwarzschild black hole  
as is the case for the metric function given by Eq.~(\ref{logistic-bet}), while for other cases
we obtain a black hole metric  as the one given by Eq.~(\ref{ferm-bet-z}). The only exception here is 
the regular black hole metric for which we have used the exponential distribution function, namely 
Eq.~(\ref{dens-exp}), for its construction.

\section{Regular black hole metric that satisfies the WEC}
\label{sec:5}

\par\noindent
In this section, we consider a distribution function different from those listed in Appendix A (and thus in Table 1).
This new distribution function is the log-logistic distribution, which is given by

\begin{equation}
\sigma(x) =  \frac{1}{(1 + x)^2}  \,\,\label{log-logist} \, .
\end{equation}

\par\noindent
As in the previous section, we will employ the metric function given in Eq.~(\ref{case-beta}). 
Thus, making the variable change $x \rightarrow 1/r$ with the appropriate factors, we can write the metric function as

\begin{equation}
f(r) = 1 - \frac{2 M}{r} \left(\frac{1}{(1+\frac{q^2}{4 \beta M r})^2}\right)^\beta
\,\,\label{log-l-bh}  \, .
\end{equation}

\par\noindent
Note that for this metric function, if we set $\beta \rightarrow 0$, then we recover the Schwarzschild black hole,  while if we set $\beta \rightarrow \infty$, then  we obtain the metric function given by Eq.~(\ref{func-exp}).

\par\noindent
It should be pointed out that the curvature invariants, namely

\bea
R &=& (2 \beta + 1)\frac{(4 \beta M)^{2\beta+1}\, q^4}{(4 \beta M r + q^2 )^{2\beta+1}} \, r^{2 \beta - 3} \,\,\label{R-sp} \, ,\\
R_{\mu\nu} R^{\mu\nu} &=& \frac{ 2^{(8\beta + 3)} (\beta M)^{4 \beta + 2} \, q^4 }{(4 \beta M r + q^2)^{4 (\beta + 1)}}   \nonumber \\
&&\times\,  [128 \beta^2 M^2 r^2 - 16 \beta (2 \beta -3) M q^2 r + (4 \beta(\beta - 1) + 5) q^4] \,
r^{4 \beta - 6}  \,\,\label{ricci-sp} \, ,\\
R_{\kappa\lambda\mu\nu} R^{\kappa\lambda\mu\nu} &=& 2^{4(2 \beta + 1)} M^2 \left(\frac{\beta^2 M^2}{(4 \beta M r + q^2)^2}\right)^{2 \beta}
[1 + \frac{( 4 \beta M r - (2 \beta -1) q^2 )^2}{(4 \beta M r + q^2)^2} \nonumber \\
&& + \frac{(16 \beta^2 M^2 r^2 - 8 \beta (2 \beta - 1) M q^2 r^2 + (\beta (2 \beta - 3) + 1)q^4)^2}{(4 \beta M r + q^2)^4}] \, r^{4 \beta - 6}
\,\,\label{krets-sp} \, ,
\eea

\par\noindent
are all regular everywhere if $\beta \geq 3/2$. 

\par\noindent
On the other hand, the WEC requires that the energy-momentum tensor must satisfy $T_{\mu\nu} t^\mu t^\nu \geq 0$
for all timelike vectors $t^\mu$~\cite{Hawking:1973uf}. In general, in terms of the distribution functions $\sigma(r)$, 
the WEC implies (see also Eqs. (8) and  (9) in Ref. \cite{Balart:2014jia})

\begin{equation}
\frac{\sigma'(r)}{r^2} \geq 0\,\,\label{wec-a} \,
\end{equation}

\par\noindent
and

\begin{equation}
2\, \frac{\sigma'(r)}{r} \geq \sigma''(r)\,\,\label{wec-b} \, .
\end{equation}

\par\noindent
It is evident that  if we set $\beta \leq 3/2$ in Eq.~(\ref{log-l-bh}) then the WEC is satisfied. 
Therefore, if we impose $\beta = 3/2$, then  we can construct a regular black hole solution which satisfies the WEC and  of which the metric function is of the form

\begin{equation}
f(r) = 1 - \frac{2 M}{r} \left(\frac{1}{1+\frac{q^2}{6 M r}}\right)^{3}\, .
\,\,\label{reg-3}  
\end{equation}

\par\noindent
In addition, this black hole solution  asymptotically behaves as the Reissner--Nordstr\"{o}m solution. 
Note that the aforesaid metric function can also be obtained as in  Ref.~\cite{Balart:2014jia}.
Moreover, it is known that if a charged regular  black hole satisfies the WEC then such a regular black hole
has de Sitter behavior at $r\rightarrow 0$. Thus, if we set $r\rightarrow 0$, then  the metric function given by Eq.  (\ref{reg-3})  behaves like

\begin{equation}
f(r) \approx 1 - 432 \frac{M^4}{q^6} r^2
\,\,\label{desitt}  \, .
\end{equation}

\par\noindent
It should be stressed that the location of the horizons, i.e., $r_\pm$, of the metric function given by Eq.  (\ref{reg-3}) can be obtained by solving the  following cubic equation:

\begin{equation}
r^3 + \left(\frac{q^2}{2 \, M} - 2 \, M \right)r^2 + \frac{q^4}{12 \, M^2} \, r + \frac{q^6}{216 \, M^3} = 0
\,\,\label{eq-reg-3}  \, .
\end{equation}

\par\noindent
It can be shown that if $q < 4 M / 3$ then Eq.~(\ref{eq-reg-3}) has two positive real roots.  
The extremal regular black hole is obtained when $q_{\mbox{\tiny{ext}}} = 4 M/3$, and then Eq.~(\ref{eq-reg-3}) 
has one positive real root of the form 

\begin{equation}
r_{h \, \mbox{\tiny{ext}}} = \frac{16 \, M}{27}
\,\,\label{r-ext}  \, .
\end{equation}

\par\noindent
Furthermore, the electric field related to metric function given by Eq.~(\ref{reg-3}) is written as

\begin{equation}
E(r) =  \frac{q}{r^2}\left(\frac{1}{1+\frac{q^2}{6 M r}}\right)^{5} \,\,\label{elec-loglog} \,
\end{equation}

\par\noindent
which asymptotically becomes that of the Maxwell theory.

\section{More regular black hole metrics that satisfy the WEC}
\label{sec:6}

\par\noindent
In this section, we will construct new black hole metrics using distribution functions 
different from those used in previous sections. These black hole metrics will depend on the arbitrary 
parameter $\beta$, and we will determine the value of $\beta$ in such a way that regularity and the WEC will be satisfied.

\par\noindent
We start by considering the standard Cauchy distribution function

\begin{equation}
\sigma(x) =  \frac{1}{(1 + x^2)}  \,\,\label{cauchy} \, 
\end{equation}

\par\noindent
and then we will make the change $x \rightarrow 1/r$ with the appropriate factors. Thus, by substituting the above distribution 
function  in Eq.~(\ref{case-beta}) and setting $\beta = 3/2$, we obtain a  black hole metric which is  regular everywhere 
and which also satisfies the WEC. It should be noted that this metric corresponds to the Bardeen solution, of which the mass function reads

\begin{equation}
m(r)=  \frac{M r^3}{(r^2 + g^2)^{3/2}} =  M \left(\frac{1}{1 + g^2/r^2}\right)^{3/2} \,\,\label{f-m-bardeen} \,
\end{equation}

\par\noindent
with $g$ being a self-gravitating magnetic monopole charge~\cite{AyonBeato:2000zs}.

\par\noindent
Next, we construct other black hole metrics based on the Dagum distribution~\cite{Dagun:1977}, which is given by

\begin{equation}
\sigma(x) =  \frac{a p x^{ap-1}}{b^{ap}(1 + (x/b)^a)^{p+1}}  \,\,\label{dagum} \, ,
\end{equation}

\par\noindent
with the variable $x > 0$ and the parameters $a, b, p > 0$.

\par\noindent
First, it is easily seen that if we choose  $a = 3$, $b = 1$ and $p = 1/3$  we can build a new metric function  
by replacing Eq. (\ref{dagum})  in Eq.~(\ref{case-beta}) and of course we have to  make the change $x \rightarrow 1/r$ 
with the appropriate factors. In this case, when we set $\beta = 3/4$, the black hole metric becomes regular
everywhere and also satisfies the WEC. This corresponds to the black hole solution given in
Ref.~\cite{Hayward:2005gi}, of which the mass function reads

\begin{equation}
m(r)=   \frac{M}{1 + 2 l^2 M/ r^3} \,\,\label{f-m-hayward} \, 
\end{equation}

\par\noindent
where $l > 0$ is the Hubble length.

\par\noindent
Second, if we choose $p = 1/a$ and $b = 1$, then we can generalize the previous case with the following distribution function:

\begin{equation}
\sigma(x) =  \frac{1}{(1 + x^a)^{\frac{a+1}{a}}}  \,\,\label{dagum-2} \, .
\end{equation}

\par\noindent
Thus, making the change $x \rightarrow q^2/(M r)$ and selecting the appropriate factors, we can write the 
black hole metric in the form

\begin{equation}
f(r) = 1 - \frac{2 M}{r} \left(\frac{1}{(1+\gamma(\frac{q^2}{M r})^a)^{\frac{a+1}{a}}}\right)^\beta
\,\,\label{dag-1-bh}  
\end{equation}

\par\noindent
where $\gamma > 0$. As before, we can calculate the curvature invariants and conclude that if we set 
$\beta \geq 3/(a + 1)$, then the black hole metric is regular everywhere; while if we set $\beta \leq 3/(a + 1)$, 
it satisfies the WEC. Thus, if we impose $\beta = 3/(a + 1)$ , we have a regular black hole metric   
which also satisfies the WEC  and is written as

\begin{equation}
f(r) = 1 - \frac{2 M}{r} \left(\frac{1}{1+\gamma \left(\frac{q^2}{M r}\right)^{a}}\right)^{3/a}   \,\,\label{f-m-dagum-gral} \, .
\end{equation}

\par\noindent
The expression of the corresponding electric field is of the form

\begin{equation}
E =\frac{q}{r^2} \left(\frac{3 \gamma (3+a)\left(\frac{q^2}{M r}\right)^{a-1}}{2\left(1+\gamma\left(\frac{q^2}{M r}\right)^{a}
\right)^{2+3/a}}\right)
\,\,\label{CE-dagum-gral} \, .
\end{equation}

\par\noindent
The metrics that can be derived  from Eq.~(\ref{f-m-dagum-gral}) asymptotically behave as the Schwarzschild black hole,  and, for small $r$, they behave as the de Sitter black hole,

\begin{equation}
f(r) \approx 1 - \frac{2 M^4}{\gamma^{3/a} q^6}r^2
\,\,\label{dag-small} \, .
\end{equation}

\par\noindent
At this point, it should be pointed out that, except for the case where $a =1$ and $\gamma=1/6$ which is equivalent to case of Eq.~(\ref{reg-3}), in all other
cases the regular black hole metrics asymptotically do not behave as the Reissner--Nordstr\"{o}m black hole. 
Furthermore, only  in the case where $a =1$ and $\gamma=1/6$, we have a model of nonlinear electrodynamics 
which, in the weak field approximation, corresponds to the Maxwell theory.

\section{Regular black holes that asymptotically behave as the Reissner--Nordstr\"{o}m solution}
\label{sec:7}

In the previous section, we  constructed a family of regular black hole metrics given by Eq.~(\ref{f-m-dagum-gral}), 
which satisfy the WEC. However, as already noted, these black hole metrics asymptotically do not behave as 
the Reissner--Nordstr\"{o}m black hole when we consider the case where $a\geq 2$. 
So, in this section, we consider two methods  to build regular black hole solutions using Eq.~(\ref{f-m-dagum-gral}), of which the electric fields will asymptotically behave as  that of the Maxwell theory. 

\par\noindent
The first method is to write the new black hole metrics in the  form

\begin{equation}
f(r) = 1 - \frac{2 M}{r} \left(\frac{1}{1+\gamma\left(\frac{q^2}{M r}\right)^{a}}\right)^{3/a}
 \frac{\sigma(r)}{\sigma(\infty)}   \,\,\label{case-prod} \,
\end{equation}

\par\noindent
where $a \geq 2$, $\gamma > 0$ is a constant and $\sigma(r)$ can be any one of  the distribution functions listed in Table 1. 
Note that the values that the charge $q$ can take in order for the black hole metric to have horizons depends on the values 
of $\gamma$.

\par\noindent
As an example of the general expression  given by Eq. (\ref{case-prod}), we consider the case in which the 
distribution function is of the form 

\begin{equation}
\sigma(r) = \frac{q^2/(M r)}{\exp\left(\frac{q^2}{M r}\right)-1} \,\,\label{dens-cas-prod} ,
\end{equation}

\par\noindent
with $a=4$ and $\gamma=1$. Thus, we obtain the following  regular black hole metric:

\begin{equation}
f(r) = 1 - \frac{2 M}{r}\left( \frac{1}{1+\left(\frac{q^{2}}{M r}\right)^4} \right)^{3/4}\left(\frac{q^2/(M r)}
{\exp\left(\frac{q^2}{M r}\right)-1}\right)\,\,\label{func-cas-2} \, .
\end{equation}

\par\noindent
In this case, the electric field associated with the aforesaid metric function is given as

\begin{equation}
E(r) = -\frac{M r^3}{2 q} \frac{d}{dr}\left(\frac{1}{r^2}\,\, \frac{d}{dr}\left(\left( \frac{1}{1+\left(\frac{q^{2}}{M r}
\right)^4} \right)^{3/4}\left(\frac{q^2/(M r)}{\exp\left(\frac{q^2}{M r}\right)-1}\right)\right)\right)\,\,\label{f-E-1} ,
\end{equation}

\par\noindent
which  corresponds to an electric field obtained from a nonlinear electrodynamics theory. This electric field is  regular everywhere and asymptotically behaves as

\begin{equation}
E(r) \approx \frac{q}{r^2} \,\,\label{f-E-1-apr} \, .
\end{equation}

\par\noindent
It is noteworthy that when $|q| = 0.907 \, M$ we obtain the metric and electric field 
for the corresponding extremal regular black hole solution.

\par\noindent
Furthermore, another interesting example which can be classified within the family of metrics described 
by Eq.~(\ref{case-prod})  was developed in Ref.~\cite{AyonBeato:1999ec}, by considering the distribution function

\begin{equation}
\sigma(r) = \exp\left(\frac{q^2}{M r}\right) \,\,\label{other-ej} \, ,
\end{equation}

\par\noindent
together with the Bardeen model, i.e., choosing $a=2$ and $\gamma=M^2/q^2$ in Eq.~(\ref{f-m-dagum-gral}).

\par\noindent
It should be stressed that all regular black hole metrics derived from the Eq.~(\ref{case-prod}) asymptotically behave 
as the Reissner--Nordstr\"{o}m black hole metric. However, these solutions do not satisfy the WEC.

\par\noindent
The second method to build more regular black hole metrics based on those given by Eq.~(\ref{f-m-dagum-gral}) 
consists of adding a new term which will make the metrics to behave asymptotically as the Reissner--Nordstr\"{o}m metric. 
To build this new term we employ once again the distribution function of Dagum with a factor $q/r^2$. 
Thus, the metric function is written

\begin{equation}
f(r) = 1 - \frac{2 M}{r} \left(\frac{1}{1+\gamma\left(\frac{q^2}{M r}\right)^{a}}\right)^{3/a} +
\frac{q^2}{r^2}\left(\frac{1}{1+\gamma\left(\frac{q^2}{M r}\right)^{a}}\right)^{4/a}  \,\,\label{BH-reg-WEC} \, ,
\end{equation}

\par\noindent
where $a \geq 2$ is an integer and $\gamma > 0$ is a constant. 
It should be pointed out that  if we set  $\gamma \geq (2/3)^a$  then the associated solution satisfies the WEC. This
can be demonstrated by calculating the corresponding distribution functions and their derivatives  
and then implementing Eqs.~(\ref{wec-a}) and~(\ref{wec-b}).
The expression for the electric field associated with the above-mentioned metric is given by

\begin{equation}
E =\frac{q}{r^2} \left(\frac{3 \gamma (3+a)\left(\frac{q^2}{M r}\right)^{a-1}}{2\left(1+\gamma\left(\frac{q^2}{M r}\right)^{a}\right)^{2+3/a}}
+ \frac{1-\gamma (3+a)\left(\frac{q^2}{M r}\right)^{a}}{\left(1+\gamma\left(\frac{q^2}{M r}\right)^{a}\right)^{2(2+a)/a}}\right)
\,\,\label{CE-reg-WEC} \, .
\end{equation}

\par\noindent
For small $r$, the black hole metrics obtained from Eq.~(\ref{BH-reg-WEC}) behave as the de Sitter black hole metric, 
namely

\begin{equation}
f(r) \approx 1 - \frac{M^4}{\gamma^{4/a} q^6}(2 \gamma^{1/a} - 1)r^2
\,\,\label{reg-WEC-small} \, .
\end{equation}

\par\noindent
Note that the factor that accompanies the term $r^2$ cannot become zero because of the restriction $\gamma \geq (2/3)^a$. 
On the other hand, if we set $(1/2)^a \leq \gamma < (2/3)^a$, the black hole metrics remain regular, without satisfying the WEC, 
but they have a de Sitter center. Therefore, if a regular black hole metric is regular and satisfies the WEC, 
then it has a  de Sitter center~\cite{Dymnikova:2001fb}. However,
if the metric has a de Sitter behavior when approaching the center, it does not necessarily satisfy the WEC. 
Furthermore,  if we set $\gamma < (1/2)^a$, the black hole metric is not regular.
\par\noindent
Finally, there is a known case which can be obtained as a particular case of Eq.~(\ref{BH-reg-WEC}) by 
choosing $a=2$ and $\gamma=M^2/q^2$.  This case corresponds to the black hole metric given in Ref.~\cite{AyonBeato:1998ub}.

\section{Conclusions}
\label{sec:8}
We have given various examples of regular black hole solutions that asymptotically behave 
as the Reissner--Nordstr\"{o}m solution, and some of them also satisfy the weak energy condition. To construct regular 
black hole metrics, we employed several probability distribution functions  in which we have replaced
the variable by its reciprocal. In addition, we have also used powers of these distribution functions as a second method 
to construct new regular black hole metrics.


It is worth noting the case based on log-logistic distribution in the Sec. 5, where we have
adjusted the power of the distribution function to get a  black hole metric which is regular everywhere and also satisfies the WEC.
One virtue of this solution is that the corresponding extremal case can be algebraically manipulated. Thus,
we can revisit the aspects studied in Refs.~\cite{Matyjasek:2004gh}, \cite{Myung:2007xd}, and~\cite{Myung:2007an} by using this solution.


\section*{Acknowledgments}

\par\noindent
L.B. would like to thank  Direcci\'on de Investigaci\'on y Postgrado de la Universidad de La Frontera  (DIUFRO).

\appendix
\section{Continuous probability distributions}\label{distrib}
Some probability distributions which satisfy $\sigma(x) > 0$ and $d\sigma(x)/dx < 0$
for $x \geq 0$ are listed here (see, e.g., Ref.~\cite{Weisstein:1999}).

\par\noindent
The exponential distribution is
\begin{equation}
\sigma(x) =  e^{-x}  \,\,\label{exp-d} \, .
\end{equation}

\par\noindent
The Fermi--Dirac distribution is
\begin{equation}
\sigma(x) =  \frac{1}{e^{x} + 1}  \,\,\label{ferm-dir} \, .
\end{equation}

\par\noindent
The logistic distribution is
\begin{equation}
\sigma(x) =  \frac{e^{-x}}{(1 + e^{-x})^2}  \,\,\label{logist} \, .
\end{equation}

\par\noindent
The hyperbolic secant distribution is~\cite{Baten:1934}
\begin{equation}
\sigma(x) =  \frac{2}{e^x + e^{-x}}  \,\,\label{sech} \, .
\end{equation}

\par\noindent
The Einstein functions are

\begin{equation}
\sigma(x) =  \frac{x}{e^x -1}  \,\,\label{einst-1} \,
\end{equation}

\par\noindent
and

\begin{equation}
\sigma(x) =  \frac{x^2 e^x}{(e^x -1)^2}  \,\,\label{einst-2} \, .
\end{equation}

\section{Dual P formalism}\label{dual}
We show here the description based in a dual representation of nonlinear electrodynamics obtained by
a Legendre transformation~\cite{Salazar:1987ap}.

\par\noindent
The metric function and its corresponding electromagnetic field arise as a solution of Einstein field equations coupled
to a nonlinear electrodynamics model; that is,
\begin{equation}
S =  \int d^4x \sqrt{-g} \left(\frac{1}{16 \pi}R - \frac{1}{4 \pi} L(F)\right) \,\,\label{action} \,
\end{equation}
where $R$ is the scalar curvature and the Lagrangian $L$ depends on $F = \frac{1}{4}F_{\mu\nu}F^{\mu\nu}$ 
which, for weak fields, describes the  Maxwell theory. 
One can describe the system under study  in terms of an auxiliary field defined by
$P_{\mu\nu} = (dL/dF) F_{\mu\nu}$. The dual representation is obtained by means of a Legendre transformation
\begin{equation}
H =  2 F \frac{d L}{d F}- L \,\,\label{Legendre} ,
\end{equation}
which is a function of the invariant $P = \frac{1}{4}P_{\mu\nu}P^{\mu\nu}$. Thus, we can express the Lagrangian $L$ depending on $P_{\mu\nu}$ as
\begin{equation}
L =  2 P \frac{d H}{d P} - H \,\,\label{Lag} \, 
\end{equation}
and the electromagnetic field as
\begin{equation}
F_{\mu\nu} = \frac{d H}{d P} P_{\mu\nu} \,\,\label{fmunu} \, .
\end{equation}
The energy-momentum tensor in the dual representation is written as
\begin{equation}
T_{\mu\nu} = \frac{1}{4 \pi}\frac{d H}{d P} P_{\mu\alpha}P_\nu^\alpha - \frac{1}{4 \pi} g_{\mu\nu} \left(2 P \frac{d H}{d P} - H \right)
\,\,\label{tensormunu} \, .
\end{equation}

\par\noindent
It follows from the components of $T_{\mu\nu}$ that $M'(r) = - r^2 H(P)$. Therefore, we can obtain the corresponding mass function.

\par\noindent
We now list the function $H(P)$ for each distribution given in Table 1, and to simplify notation
we define $U = s\sqrt[4]{-2 q^2 P}$, $s = q/(2 M)$.\\

\par\noindent
Exponential:
\begin{equation}
H = P e^{-U} \,\,\label{H-exp} \, 
\end{equation}

\par\noindent
Fermi--Dirac:
\begin{equation}
H = 4 P \frac{e^{2 U}}{(1 + e^{2 U})^2} = P(1-\tanh^2(U))
\,\,\label{H-fermi-dirac} 
\end{equation}

\par\noindent
Logistic:
\begin{equation}
H = \frac{P}{\sqrt{U}} \mbox{sech}^2 \sqrt{U} \tanh \sqrt{U}
\,\,\label{H-logist} \, 
\end{equation}

\par\noindent
Hyperbolic secant:
\begin{equation}
H = \frac{P}{\sqrt{2 U}} \mbox{sech} \sqrt{2 U} \tanh \sqrt{2 U}
\,\,\label{H-hip-sec} \, 
\end{equation}

\par\noindent
Einstein:
\begin{equation}
H = P (4 U e^{2 U} -2) \frac{1}{e^{2 U} - 1}
\,\,\label{H-einst-1} \, ,
\end{equation}
\begin{equation}
H = 3 P \, \mbox{csch}^2(\sqrt{3 U})\left(\sqrt{3 U} \coth(\sqrt{3 U})-1\right)
\,\,\label{H-einst-2} \, .
\end{equation}

\par\noindent
Finally, we also include the case inspired by the log-logistic distribution which reads
\begin{equation}
H = \frac{P}{\left(1 + U/3\right)^4}
\,\,\label{H-log-logist} \, .
\end{equation}


\begin{thebibliography}{}




\bibitem{Ansoldi:2008jw}
  S.~Ansoldi,
  arXiv:0802.0330 [gr-qc].

\bibitem{Elizalde:2002yz}
  E.~Elizalde and S.~R.~Hildebrandt,
  Phys.\ Rev.\  D {\bf 65}, 124024 (2002).

\bibitem{Zaslavskii:2010qz}
  O.~B.~Zaslavskii,
  Phys.\ Lett.\  B {\bf 688}, 278 (2010).

\bibitem{Dymnikova:2001fb}
  I.~Dymnikova,
  Class.\ Quant.\ Grav.\  {\bf 19}, 725 (2002).

\bibitem{Breton:2004qa}
  N.~Breton,
  Gen.\ Rel.\ Grav.\  {\bf 37}, 643 (2005).

\bibitem{Bose:1998uu}
  S.~Bose and N.~Dadhich,
  Phys.\ Rev.\  D {\bf 60}, 064010 (1999).

\bibitem{Balart:2009xr}
  L.~Balart,
  Phys.\ Lett.\  B {\bf 687}, 280 (2010).

\bibitem{AyonBeato:1998ub}
  E.~Ayon-Beato and A.~Garcia,
  Phys.\ Rev.\ Lett.\  {\bf 80}, 5056 (1998).

\bibitem{AyonBeato:2004ih}
  E.~Ayon-Beato and A.~Garcia,
  Gen.\ Rel.\ Grav.\  {\bf 37}, 635 (2005).

\bibitem{Dymnikova:2004zc}
  I.~Dymnikova,
  Class.\ Quant.\ Grav.\  {\bf 21}, 4417 (2004).

\bibitem{AyonBeato:1999rg}
  E.~Ayon-Beato and A.~Garcia,
  Phys.\ Lett.\  B {\bf 464}, 25 (1999).

\bibitem{AyonBeato:1999ec}
  E.~Ayon-Beato and A.~Garcia,
  Gen.\ Rel.\ Grav.\  {\bf 31}, 629 (1999).

\bibitem{Bronnikov:2000vy}
  K.~A.~Bronnikov,
  Phys.\ Rev.\  D {\bf 63}, 044005 (2001).

\bibitem{Bardeen:1968}
J. M. Bardeen, in Proceedings of GR5, Tbilisi, U.S.S.R, p.174 (1968).

\bibitem{AyonBeato:2000zs}
  E.~Ayon-Beato and A.~Garcia,
  Phys.\ Lett.\  B {\bf 493}, 149 (2000).

\bibitem{Balart:2014jia} 
  L.~Balart and E.~C.~Vagenas,
  Phys.\ Lett.\ B {\bf 730}, 14 (2014).

\bibitem{Nicolini:2005vd}
  P.~Nicolini, A.~Smailagic and E.~Spallucci,
  Phys.\ Lett.\  B {\bf 632}, 547 (2006).

\bibitem{Ansoldi:2006vg}
  S.~Ansoldi, P.~Nicolini, A.~Smailagic and E.~Spallucci,
  Phys.\ Lett.\  B {\bf 645}, 261 (2007).

\bibitem{Dymnikova:1992ux}
  I.~Dymnikova,
  Gen.\ Rel.\ Grav.\  {\bf 24}, 235 (1992).

\bibitem{Hayward:2005gi}
  S.~A.~Hayward,
  Phys.\ Rev.\ Lett.\  {\bf 96}, 031103 (2006).

\bibitem{Lemos:2011dq}
  J.~P.~S.~Lemos and V.~T.~Zanchin,
  Phys.\ Rev.\  D {\bf 83}, 124005 (2011).

\bibitem{Eiroa:2010wm}
  E.~F.~Eiroa and C.~M.~Sendra,
  Class.\ Quant.\ Grav.\  {\bf 28}, 085008 (2011).

\bibitem{Myung:2007qt}
  Y.~S.~Myung, Y.~W.~Kim and Y.~J.~Park,
  Phys.\ Lett.\  B {\bf 656}, 221 (2007).

\bibitem{Mo:2006tb}
  W.~J.~Mo, R.~G.~Cai and R.~K.~Su,
  Commun.\ Theor.\ Phys.\  {\bf 46}, 453 (2006).

\bibitem{Myung:2007av}
  Y.~S.~Myung, Y.~W.~Kim and Y.~J.~Park,
  Gen.\ Rel.\ Grav.\  {\bf 41}, 1051 (2009).

\bibitem{Matyjasek:2004gh}
  J.~Matyjasek,
 Phys.\ Rev.\  D {\bf 70},  047504  (2004).

\bibitem{Myung:2007xd}
  Y.~S.~Myung, Y.~W.~Kim and Y.~J.~Park,
  Phys.\ Lett.\  B {\bf 659}, 832  (2008).

\bibitem{Myung:2007an}
  Y.~S.~Myung, Y.~W.~Kim and Y.~J.~Park,
  Phys.\ Rev.\  D {\bf 76}, 104045  (2007).

\bibitem{Matyjasek:2008kn}
  J.~Matyjasek, D.~Tryniecki and M.~Klimek,
  Mod.\ Phys.\ Lett.\  A {\bf 23}, 3377  (2008).

\bibitem{Matyjasek:2000iy}
  J.~Matyjasek,
 Phys.\ Rev.\  D {\bf 63}, 084004  (2001).

\bibitem{Berej:2002xd}
W.~Berej and J.~Matyjasek,
 Phys.\ Rev.\  D {\bf 66},  024022  (2002).

\bibitem{Matyjasek:2007zz}
  J.~Matyjasek,
  Phys.\ Rev.\  D {\bf 76},  084003  (2007).

\bibitem{Berej:2006cc}
  W.~Berej, J.~Matyjasek, D.~Tryniecki and M.~Woronowicz,
  Gen.\ Rel.\ Grav.\  {\bf 38},  885  (2006).

\bibitem{Matyjasek:2008yq}
  J.~Matyjasek,
  Acta Phys.\ Polon.\  B {\bf 39},  3  (2008).

\bibitem{Lobo:2006xt}
  F.~S.~N.~Lobo and A.~V.~B.~Arellano,
  Class.\ Quant.\ Grav.\  {\bf 24},  1069  (2007).

\bibitem{Hollenstein:2008hp}
  L.~Hollenstein and F.~S.~N.~Lobo,
Phys.\ Rev.\  D {\bf 78}, 124007  (2008).

\bibitem{Moreno:2002gg}
  C.~Moreno and O.~Sarbach,
  Phys.\ Rev.\  D {\bf 67}, 024028 (2003).

\bibitem{Breton:2005ye}
  N.~Breton,
  Phys.\ Rev.\  D {\bf 72}, 044015 (2005).

\bibitem{Breton:2014nba} 
  N.~Breton and S.~E.~P.~Bergliaffa,
  arXiv:1402.2922 [gr-qc].

\bibitem{Salazar:1987ap}
  I.~H.~Salazar, A.~Garcia and J.~Plebanski,
 J.\ Math.\ Phys.\  {\bf 28}, 2171  (1987).
  
  
\bibitem{Hawking:1973uf}
  S.~W.~Hawking and G.~F.~R.~Ellis,
  {\it The Large Scale Structure of Space-Time} 
(Cambridge University Press, Cambridge, England, 1973).

\bibitem{Dagun:1977}
C.~Dagun,
Economie Appliqu\'ee. {\bf 30}, 413 (1977).

\bibitem{Weisstein:1999}
  E.~W.~Weisstein,
  {\it CRC Concise Encyclopedia of Mathematics} 
(CRC Press, Boca Raton, Florida, 1999).


\bibitem{Baten:1934}
W.~D.~Baten,
Bull. Amer. Math. Soc. {\bf 40}, 284 (1934).

\end{thebibliography}
%

\end{document}